\documentclass{article}
\usepackage{arxiv}

\usepackage[utf8]{inputenc} 
\usepackage[T1]{fontenc}    
\usepackage{hyperref}       
\usepackage{url}            
\usepackage{booktabs}       
\usepackage[dvipsnames]{xcolor}
\usepackage{xcolor}
\usepackage{colortbl}
\usepackage{amsfonts}       
\usepackage{amsmath}        
\usepackage{bm}             
\usepackage{microtype}      
\usepackage{graphicx}
\usepackage{authblk}
\usepackage{setspace}
\usepackage{stackengine}
\newcommand\xrowht[2][0]{\addstackgap[.5\dimexpr#2\relax]{\vphantom{#1}}}
\title{Deep Learning Level-3 Electron Trigger for CLAS12}

\author[1]{R.~Tyson}
\author[2,*]{G.~Gavalian}
\author[1]{D.~G.~Ireland}
\author[1]{B.~McKinnon}
\affil[1]{SUPA, School of Physics and Astronomy, University of Glasgow, Glasgow G12 8QQ, United Kingdom}
\affil[2]{Jefferson Lab, Newport News, VA, USA}
\affil[*]{Corresponding author: {gavalian@jlab.org}}

\begin{document}

\maketitle

\begin{abstract}
Fast, efficient and accurate triggers are a critical requirement for modern high energy physics experiments given the increasingly large quantities of data that they produce. The CEBAF Large Acceptance Spectrometer (CLAS12) employs a highly efficient electron trigger to filter the amount of recorded data by requiring at least one electron in each event, at the cost of a low purity in electron identification. Machine learning algorithms are increasingly employed for classification tasks such as particle identification due to their high accuracy and fast processing times. In this article we show how a convolutional neural network could be deployed as a Level 3 electron trigger at CLAS12. We demonstrate that the AI trigger would achieve a significant data reduction compared to the traditional trigger, whilst preserving a 99.5\% electron identification efficiency. The AI trigger purity as a function of increased luminosity is improved relative to the traditional trigger. As a consequence, this AI trigger can achieve a data recording reduction improvement of 0.33\% per nA when compared to the traditional trigger whilst maintaining an efficiency above 99.5\%. A reduction in data output also reduces storage costs and post-processing times, which in turn reduces time to publication of new physics measurements. 

\end{abstract}

\section{Introduction}

\paragraph{}
The Continuous Electron Beam Accelerator Facility (CEBAF) produces and delivers an electron beam to the four experimental halls of the Thomas Jefferson National Accelerator Facility (JLab), including the CLAS12 (CEBAF Large Acceptance Spectrometer @ 12 GeV) detector \cite{C12Overview} located in Hall B. The CLAS12 experimental program broadly encompasses electroproduction experiments aiming to further the global understanding of hadronic structure and Quantum Chromodynamics \cite{C12Program}. The CLAS12 detector was built to have full azimuthal angular coverage and a large acceptance in polar angle, allowing measurements to be made over large kinematic ranges \cite{C12Overview}. Very low polar angular coverage, from 2.5 to 5 degrees, is enabled by the forward tagger \cite{C12FT}, whilst the forward detector covers the range of polar angles from 5 to 35 degrees and is segmented into six sectors of azimuthal angle. The central detector covers the range of polar angles from 35 to 125 degrees, but as it was not designed to detect electrons, we can ignore it for the remainder of this article. Given the electroproduction nature of its experiments, a critical aspect of data taking using CLAS12 is the electron trigger. This trigger flags events with at least one detected electron, therefore allowing the filtering of data recorded to tape which in turn reduces storage costs and post processing times.

\begin{figure}[ht!]
\centering 
\includegraphics[width=0.5\textwidth]{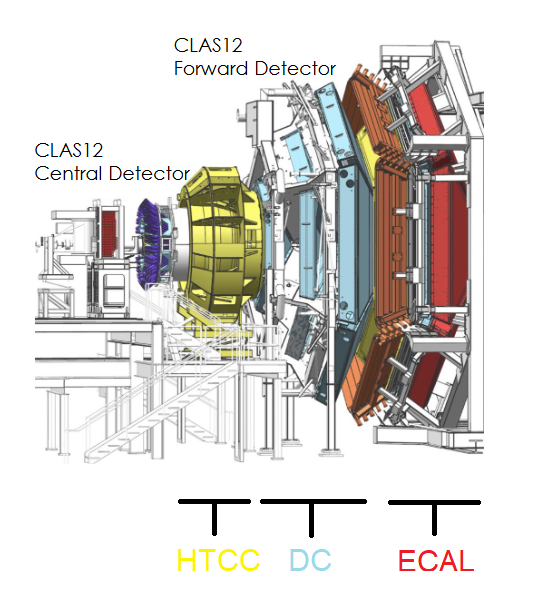}
\caption{A diagram of the CLAS12 detector showing the central and forward detectors and highlighting relevant subsystems of the forward detector used to identify electrons, as discussed in Section 2. Three of the forward detector's six sectors are visible. The forward tagger covering low polar angles is here hidden from view and is situated close to the beam-line. Mentioned for completeness, the central detector is not used to detect electrons as it does not have the capabilities to identify them and will therefore be ignored for the remainder of this article.}
\label{fig:C12_DET}
\end{figure}

\paragraph{}
The CLAS12 trigger system \cite{C12Trigger} was designed to efficiently select events for the different experiments comprising the CLAS12 physics program. Specifically, it selects events with a scattered electrons detected in the CLAS12 forward detector or forward tagger. The forward tagger is composed of a hodoscope which discriminates between electrons and photons, and a calorimeter which measures the energy deposited by individual particles. The electron trigger in the forward tagger requires a coincidence of a hit in the hodoscope and a cluster with energy within a specified range \cite{C12Trigger}. Identifying electrons, on which to form a trigger, in the forward detector is more complicated since the forward detector detects many more particle types. Resolving these different particle types requires more sophisticated particle identification algorithms, as will be discussed in Section 2. A diagram of the CLAS12 detector is shown in figure ~\ref{fig:C12_DET}, with the forward tagger here hidden from view.

\paragraph{}
The performance of the trigger can be evaluated based on two metrics, the purity and the efficiency. The easiest way to describe these is by using a confusion matrix, as shown in table ~\ref{table:ConfMat}. A True Positive (TP) is an event with an electron in a given sector that was correctly selected by the trigger. A False Negative (FN) is an event with an electron in a given sector that was incorrectly rejected by the trigger. A True Negative (TN) is an event with no electrons in a given sector that was correctly rejected by the trigger. A False Positive (FP) is an event with no electrons in a given sector that was incorrectly selected by the trigger. This can happen when the trigger misidentifies another particle type, or even a spurious hit, as an electron.

\paragraph{}
The efficiency (E), purity (P) and overall accuracy (A) of the trigger are then calculated as:
\begin{equation} E=\frac{TP}{TP+FN} \end{equation}
\\
\begin{equation} P=\frac{TP}{TP+FP} \end{equation}
\\
\begin{equation} A=\frac{TP+TN}{TP+FP+FN+TN} \end{equation}

\begin{table}[h!]
\centering
\begin{tabular}{ |c|c|c| } 

 \hline
 \xrowht[()]{30pt}
 \cellcolor{white} Confusion Matrix & \cellcolor{blue!40} Electron in Sector & \cellcolor{red!60} No Electron in Sector\\
 \hline\xrowht{30pt}
 \cellcolor{blue!40} Selected by Trigger & \cellcolor{blue!40} True Positive (TP) & \cellcolor{violet!60} False Positive (FP)\\
 \hline\xrowht{30pt}
\cellcolor{red!60} Rejected by Trigger & \cellcolor{violet!60} False Negative (FN) & \cellcolor{red!60} True Negative (TN)\\
 \hline
\end{tabular}
\caption{The electron trigger confusion matrix.}
\label{table:ConfMat}
\end{table}

\paragraph{}
Despite achieving high efficiency in selecting events with at least one electron in at least one of the six sectors of the forward detector \cite{C12Trigger}, the CLAS12 electron trigger can misclassify events where another particle type is mistaken for the electron in one of the six sectors. This decreases the purity of the trigger. Typically a balancing game can be played between the purity and efficiency in which as one decreases the other increases, therefore easing or tightening the requirements on electron identification. This is reflected by the overall accuracy which is highest when maximising the product of the purity and efficiency.

\paragraph{}
Machine learning algorithms are known to be very effective for classification tasks and previous research has demonstrated their use as triggers. In \cite{Lit1}, it was shown that Multi Layer Perceptrons could use low level information from photomultiplier tubes at the Jiangmen Underground Neutrino observatory (JUNO) to select neutrino events with a high efficiency whilst decreasing the required logic ressources. Similarly, \cite{Lit2} demonstrates how neural networks could process low level information in Drift Tube chambers at the Compact Muon Solenoid (CMS) to improve muon identification for triggering by using a network to remove noise from the Drift Tubes and a separate network to identify the side of passage of the hit with respect to a wire within the Drift Tube. Both of these examples constitute so called Level 1 triggers that take raw hits from a single type of detectors to make a trigger decision. The LHCb collaboration also employs machine learning algorithms such as boosted decision trees for High Level triggers \cite{Lit3}. These are typically based on variables reconstructed from post processing of low level information, with the aim of identifying specific channels or reactions such as $B$ production.

\paragraph{}
In this article we present an AI electron trigger for CLAS12 capable of higher purity electron selection than the traditional electron trigger. Our AI trigger employs a convolutional neural network applied to low level information taken from hits in two different detectors. The trigger we obtain in this way sits somewhere in between the two kinds of triggers cited above: we construct a Level 3 trigger which aims to identify electrons by combining low level information from various CLAS12 subsystems. As such, the trigger can run online without requiring any data post processing whilst being able to better identify electrons than the traditional CLAS12 trigger.

\section{CLAS12 Electron Trigger}

\paragraph{}
The CLAS12 Electron trigger \cite{C12Trigger} uses information from the forward Electromagnetic Calorimeter \cite{C12ECAL} (ECAL), the drift chambers \cite{C12DC} (DC) and the High Threshold Cherenkov Counter \cite{C12HTCC} (HTCC) to select events with an electron in at least one of the six CLAS12 Forward Detector sectors.

\begin{figure}[ht!]
\centering  
\includegraphics[width=0.45\textwidth]{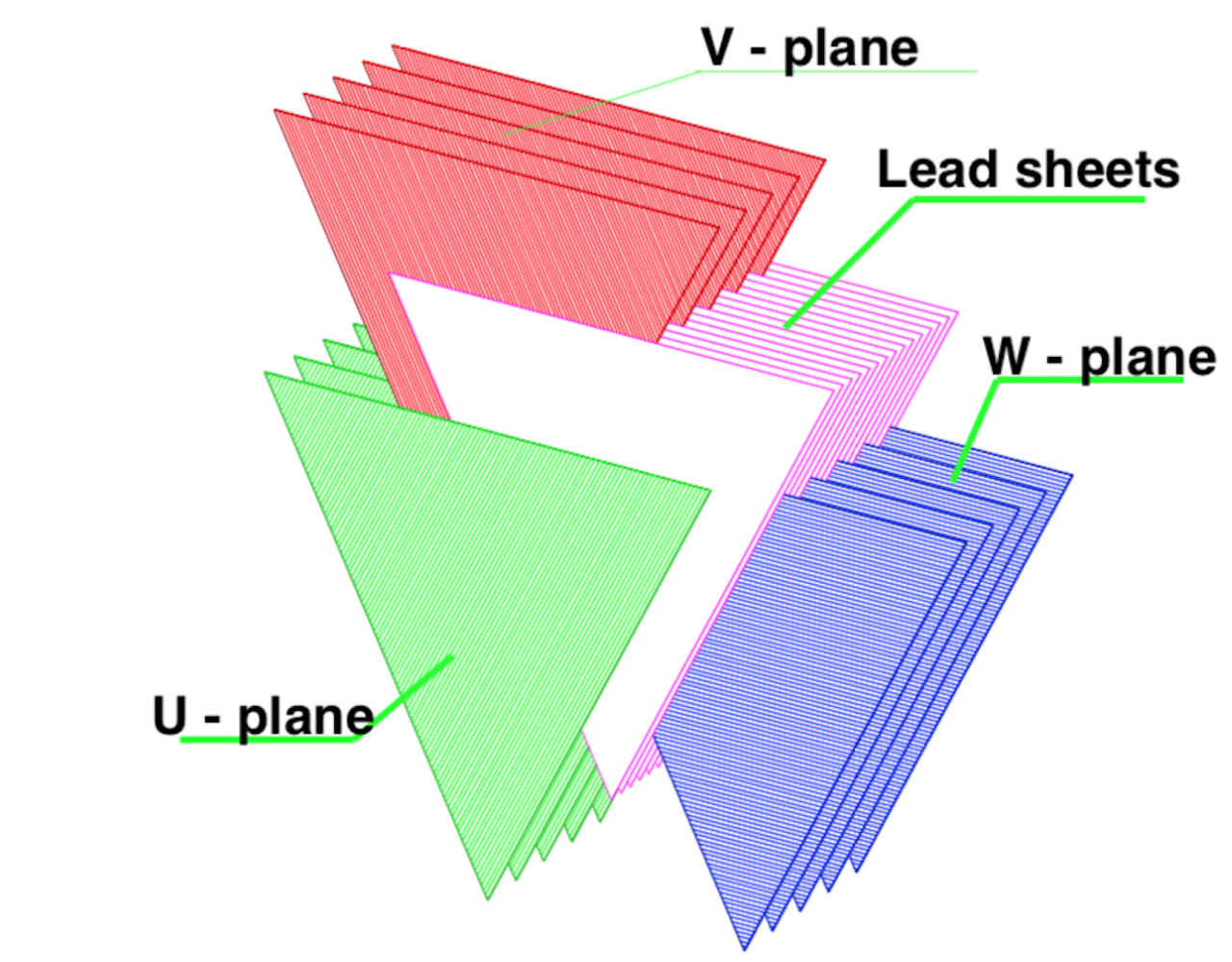}
\includegraphics[width=0.45\textwidth]{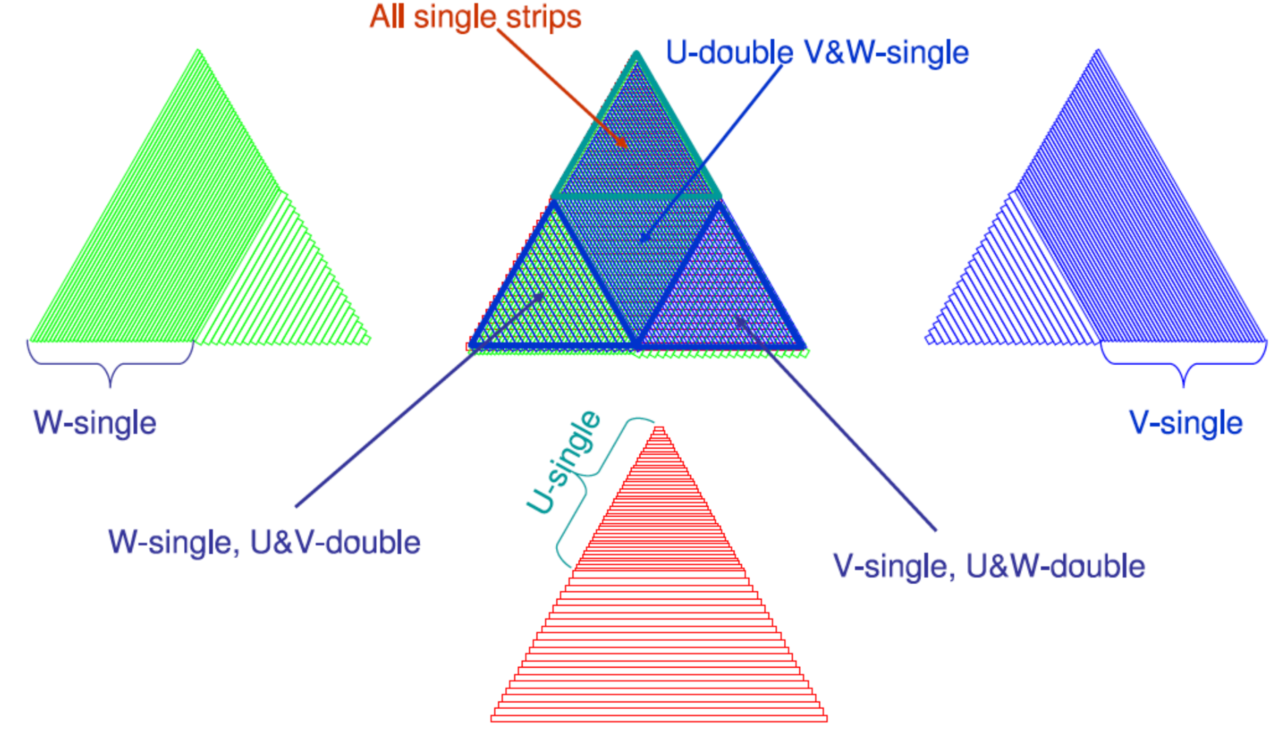}
\caption{Schematic representation of the calorimeters' three views taken from \cite{C12ECAL}.}
\label{fig:ECALDiag}
\end{figure}

\paragraph{}
The ECAL is composed of two separate subsystems, the electromagnetic calorimeter (EC) and the pre-shower calorimeter (PCAL), both composed of three views (U/V/W), as exemplified in figure~\ref{fig:ECALDiag}. The EC itself contains two calorimeters, the EC inner and EC outer, which have 36 strips in each of their three views. The PCAL has 68 strips in the U view and 62 in V and W \cite{C12ECAL}. The primary purpose of these detectors is electron identification via the energy of their electromagnetic showers as electrons will deposit more energy than hadrons in the ECAL \cite{C12Trigger}. 

\paragraph{}
The HTCC was specially designed to discriminate electrons from other charged particles. The HTCC consists of 60 mirror sections readout by 48 photo-multiplier tubes (PMTs) with clusters made of one to four PMT signals collecting the Cherenkov light from adjacent mirrors \cite{C12HTCC}. The number of photo-electrons produced in a cluster then becomes a discriminating variable in identifying electrons, with the requirement being at least two photo-electrons produced in the HTCC for electrons \cite{C12Trigger}.

\begin{figure}[ht!]
\centering  
\includegraphics[width=0.75\textwidth]{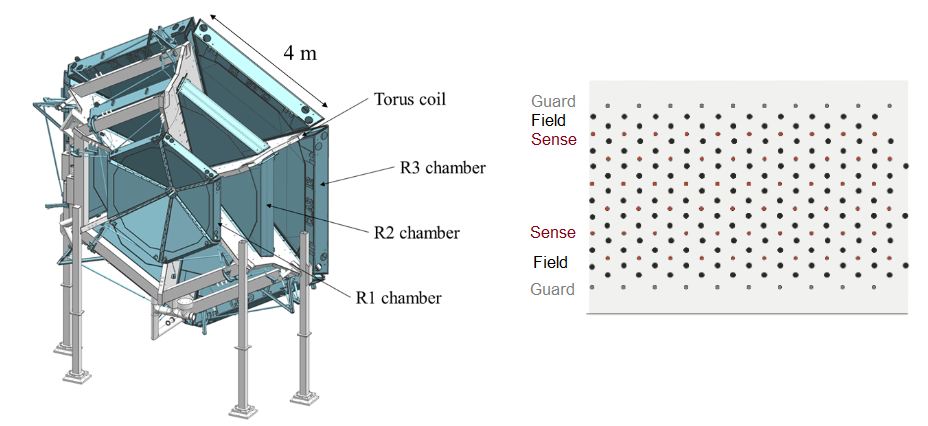}
\caption{Left: Diagram showing the three chambers of the DC, each chamber containing two superlayers. The torus magnet allows to measure the charge and momentum of charged particles travelling through the DC. The forward detector's six sectors are also visible. Right: Schematic of the wire layout for one superlayer, with each superlayer containing 6 layers of sense wires (red). The view is a cut perpendicular to the wire direction. Both images taken from \cite{C12DC}.}
\label{fig:DCDiag}
\end{figure}

\paragraph{}
The DC, shown in figure~\ref{fig:DCDiag} is made of six superlayers in each of the six CLAS12 sectors, with six layers per superlayer and 112 wires per layer \cite{C12DC}. The trigger decision requires at least three layers in every superlayer and at least five superlayers in every sector. There is no signal amplitude information available therefore only hit information can be used by the trigger \cite{C12Trigger}.

\begin{figure}[ht!]
\centering  
\includegraphics[width=0.5\textwidth]{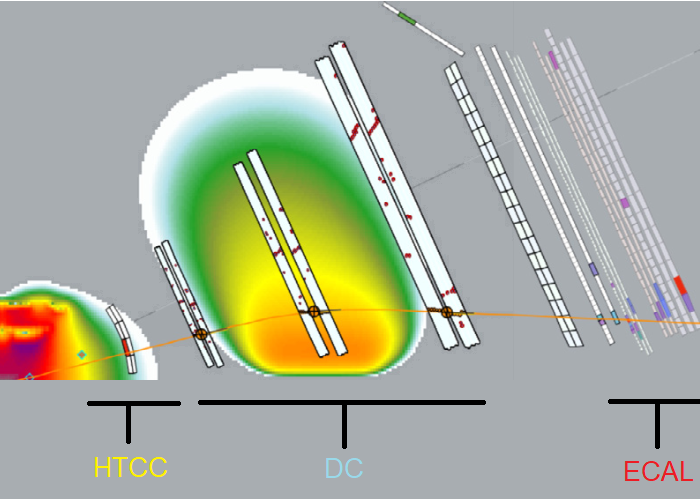}
\caption{A diagram of an electron track reconstructed in the forward detector, adapted from \cite{C12Overview}. The heat maps represent the magnetic field strength close to the target and within the forward detector.}
\label{fig:CED}
\end{figure}

\paragraph{}
The electron trigger then combines information from all three of these detectors to select events with at least one electron in one of the six sectors. This is done in parallel for all sectors by requiring at least two photoelectrons produced in the HTCC and high energy deposition in the calorimeters \cite{C12Trigger}. Geometrical matching between the HTCC signal, the PCAL shower and track candidates in the DC helps to further remove background events without a scattered electron in one of the six sectors \cite{C12Trigger}. Figure ~\ref{fig:CED}, shows an electron going through the various subsystems of the CLAS12 detector where the red line here denotes an electron track as reconstructed by the CLAS12 data post-processing \cite{C12Software}.

\begin{figure}[ht!]
\centering  
\includegraphics[width=0.65\textwidth]{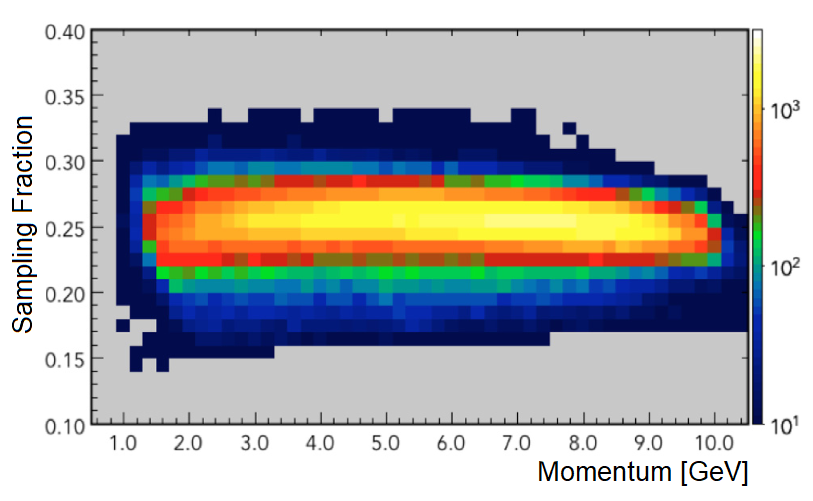}
\caption{The electron sampling fraction at CLAS12.}
\label{fig:SF}
\end{figure}

\paragraph{}
As mentioned earlier, the performance of the electron trigger can then be measured using several metrics. The efficiency measures the proportion of events with an electron which were selected by the trigger. The efficiency is hard to measure offline given that by definition the events with at least one electron that were incorrectly rejected by the trigger won't be saved. Instead, to compare to the work presented here, we refer to the reported measurements placing the efficiency over 99.5\% in all electron momentum bins \cite{C12Trigger}. For this work, the purity of the electron trigger was measured as the proportion of events accepted by the trigger in a given sector with a particle reconstructed in the same sector that was identified as an electron by the usual CLAS12 data post-processing \cite{C12Software}. This improves on the simple criteria required by the trigger first with a more rigorous matching of tracks associated with the electron in the DC to clusters in the ECAL and HTCC, and other CLAS12 subsystems not mentioned here for brevity. The same requirements are made on the charge of the particle, the number of photoelectrons in the HTCC and the lower threshold on the energy deposition in the ECAL. However, an additional requirement is made on the electron sampling fraction, defined as the ratio of the energy deposition to the momentum of a particle. As seen in figure ~\ref{fig:SF}, the sampling fraction for electrons is centered around 0.25. A parametrisation is established for both the mean and standard deviation of a normal distribution fitted to slices of the sampling fraction in bins of momentum. The sampling fraction for electrons is then required to be within five standard deviations from the mean for a given momentum. The two additional requirements on track matching and sampling fraction remove most other charged particle types, greatly increasing the particle identification purity, but require offline processing and cannot be used for the online trigger.

\paragraph{}
The trigger purity will change depending on the running conditions. At higher luminosities, the resulting higher occupancy means the purity decreases. This happens, for example, by increasing the beam current. The data set used for this article was taken with a liquid hydrogen target ($LH_{2}$) and a 10.6 GeV electron beam. The purity at different beam currents is shown in table ~\ref{table:Purity}.

\begin{table}[h!]
\centering
\begin{tabular}{ |c||c| } 
 \hline
 Beam Current & Purity \\ 
 \hline
 \hline
 5 nA & 43\% \\
 \hline
 40 nA & 28\%  \\
 \hline
 45 nA & 29\%  \\
 \hline
 50 nA & 27\%  \\
 \hline
 55 nA & 23\%  \\
 \hline
\end{tabular}
\caption{The CLAS12 electron trigger purity measured at different beam currents.}
\label{table:Purity}
\end{table}

\begin{figure}[ht!]
\centering  
\includegraphics[width=0.99\textwidth]{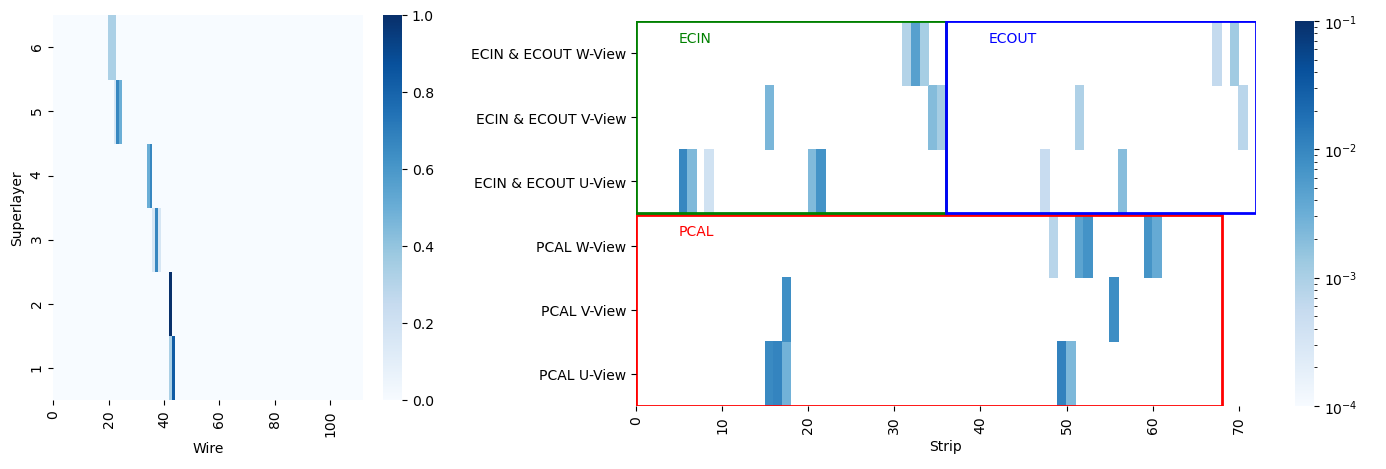}
\caption{Left: An example of a DC image containing an electron track in a given sector. The colour axis represents the number of layers with the same wire number, divided by six as normalisation. Right: an example of an ECAL image containing all ECAL hits in a given sector for a single event. The colour axis represents the energy deposited in each strip divided by three.}
\label{fig:Feature}
\end{figure}

\paragraph{}
The aim of the AI Trigger is to improve the purity. To this end, hits in the DC and the ECAL were composed into separate images from which the machine learning classifier will be able to learn the energy and momentum distributions characteristic of electrons. The DC images are effectively 6$\times$112 arrays where the rows represent each of the six superlayers and the columns represent the wires. These arrays are then filled with 0 if a wire did not record a hit or with 1/6 per hit per layer. The hits are taken on a sector per sector basis, as the AI trigger will be applied on individual sectors, meaning that there can be hits from multiple tracks in each of the individual DC images. The ECAL images are effectively 6$\times$72 arrays where the rows represent the U/V/W view for the PCAL then the EC inner and EC outer arrays which are stacked side by side on the last three rows. The columns represent the individual strips. These arrays are then filled with the energy deposited in each of the strips, divided by three as three GeV was taken as the upper limit on the energy than can be deposited in an individual strip. No further normalisation is then applied to either the DC or ECAL images. An example of both is given ~\ref{fig:Feature}.

\begin{figure}[ht!]
\centering  
\includegraphics[width=0.99\textwidth]{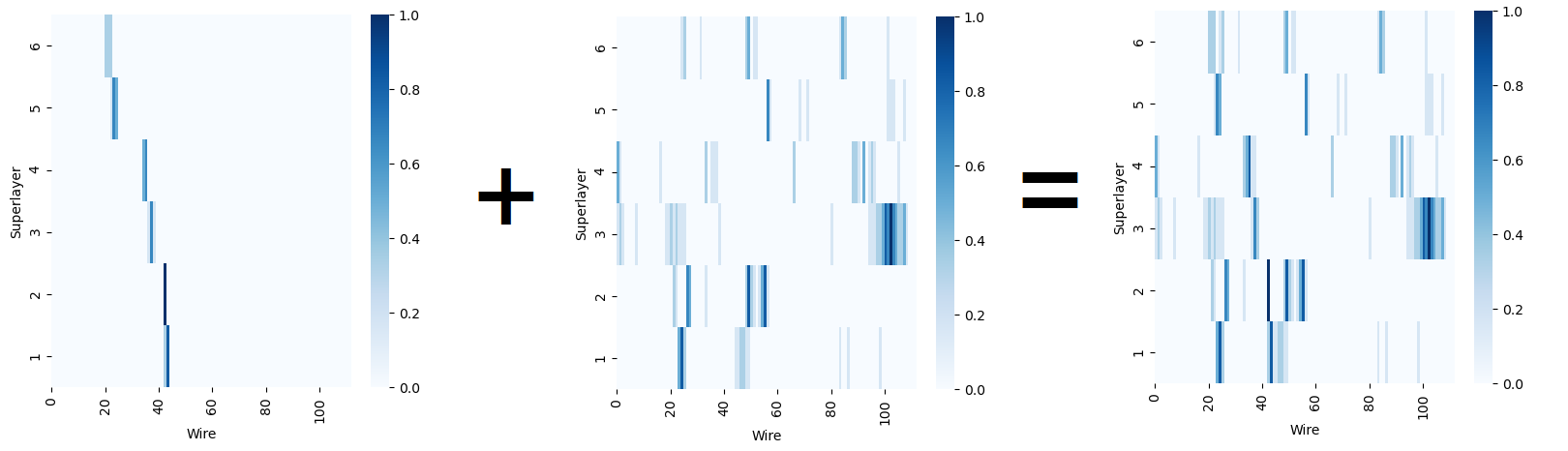}\\
\includegraphics[width=0.99\textwidth]{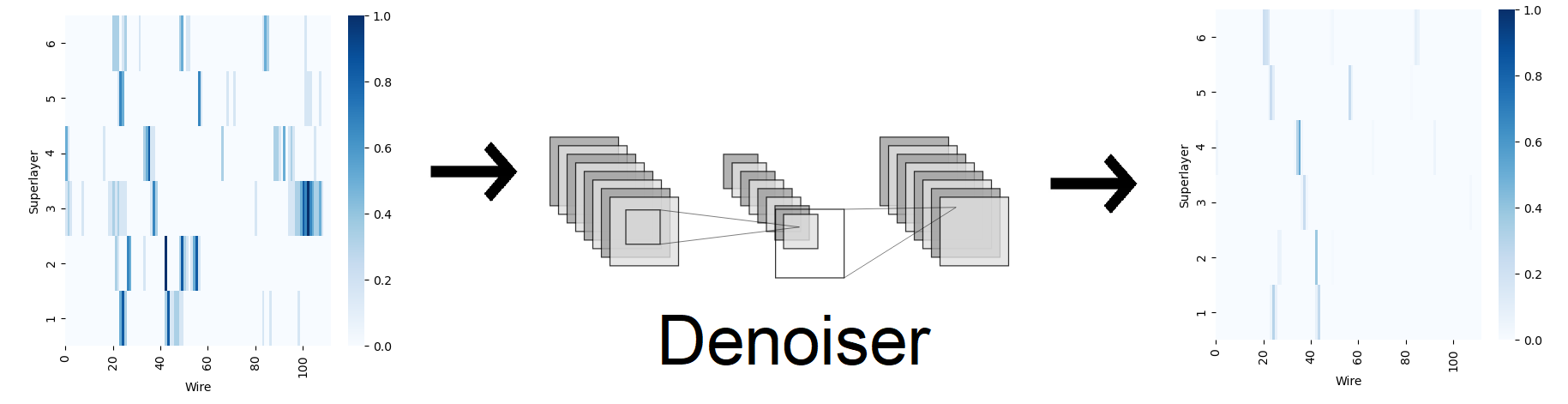}
\caption{The top row demonstrates the process of adding spurious noise to a clean track to synthetically create raw hits as measured by the DC. The bottom row demonstrates how the de-noiser can be applied to the raw hits to return a cleaner track.}
\label{fig:DCNoise}
\end{figure}

\paragraph{}
The example in figure ~\ref{fig:Feature} shows a clean electron track in the DC. In practice however, the raw hits in the DC will contain additional spurious hits, as shown in the first row of figure ~\ref{fig:DCNoise}. This can essentially be thought of as noise and we sought to investigate whether this would have an impact on the AI trigger performance. To this end, we take the clean DC tracks as reconstructed by the usual CLAS12 reconstruction software, to which we add background taken at 45 nA, 50 nA, and 55 nA to synthetically recreate the raw hits measured in the DC at these various beam currents. The background is typically taken from random trigger data which was initially identified to emulate physics and electronic backgrounds in the various detectors when simulating the CLAS12 detector \cite{GEMC}. Additionally, we can double the background merged with the clean tracks to simulate data taken at 90 nA, 100 nA and 110 nA \cite{GEMC}. From there we can train on the raw hits simulated with the background merging. Alternatively, we can follow the approach described in \cite{denoise} to de-noise the DC tracks. This allows for a realistic comparison between applying the trigger on raw hits or clean tracks in the DC. Both approaches could be deployed during data taking. To train the de-noising algorithm we followed the procedure described in \cite{denoise}, modifying the proposed architecture only by simplifying it in the interest of achieving higher prediction rates. The output of the de-noiser can be found in the bottom row of figure ~\ref{fig:DCNoise}, showing much cleaner tracks that resemble that shown in figure ~\ref{fig:Feature}.

\section{Training}
\begin{figure}[ht!]
\centering  
\includegraphics[width=0.75\textwidth]{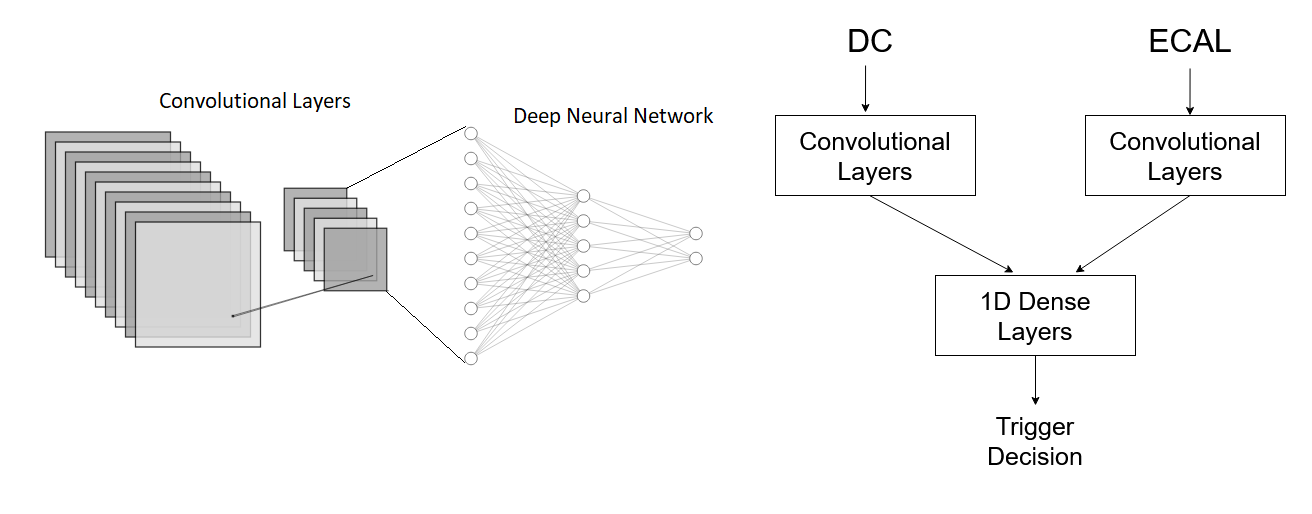}
\caption{Left: an example of a 2D Convolutional Neural Network. Right: the architecture chosen for the AI trigger CNN. The DC and ECAL images are fed to separate sets of convolutional layers.}
\label{fig:CNNDiag}
\end{figure}

\paragraph{}
Convolutional neural networks \cite{CNN1,CNN2} (CNNs) have become the go-to neural networks for image classification and more generally computer vision. These are composed of convolutional layers, which enhance or remove features of their input images, and a deep neural network which classifies the output of the convolutional layers, as shown in figure ~\ref{fig:CNNDiag}. For this project we chose a specific architecture where the DC and ECAL images are passed to separate sets of convolutional layers. The output of both sets of convolutional layers are recast into 1D arrays and concatenated into a single array which is then fed to a deep neural network for classification. The logic behind this choice is that such an architecture leaves open the possibility of adding information from  additional detectors without affecting the structure of the convolutional layers investigated here. The rightmost figure of ~\ref{fig:CNNDiag} shows a diagram of the AI trigger architecture.

\paragraph{}
As a CNN trains, each of the filters will be optimised, along with the weights of the deep neural network, so as to improve the CNN's performance. Several hyperparameters can then be tuned by the user, such as the number of nodes and layers in the neural network, the number of convolutional layers and filters per layer, and the stride which represents the step size as the filter is scanned across the image. Here, training was done in python with the TensorFlow library \cite{tf}  using the Adam optimiser at a learning rate of $1\times10^{-5}$, and a negative log likelihood loss function. Whilst optimising the network architecture, we found that varying the number of layers and nodes per layer in the deep neural network did not have a major impact on the purity or efficiency of the AI trigger, and decided to use three layers with 1000, 500, and 2 nodes respectively. Next we tuned the number of convolutional layers and filters in each of these. It was found that adding more filters per layer increased the purity at an efficiency above 99.5\% but had little impact on the prediction time as long as the number of filters did not increase too much past 64. On the contrary, adding more than two layers slowed the CNN down without improving its performance, although adding a second layer did provide a marked improvement in performance. The chosen architecture kept the two branches for the DC and ECAL images identical, each having two convolutional layers with 64 and 16 filters respectively. The stride of the filters was set to (1,2) to decrease the number of operations performed by the CNN and increase its prediction rate.

\begin{figure}[ht!]
\centering 
\includegraphics[width=0.45\textwidth]{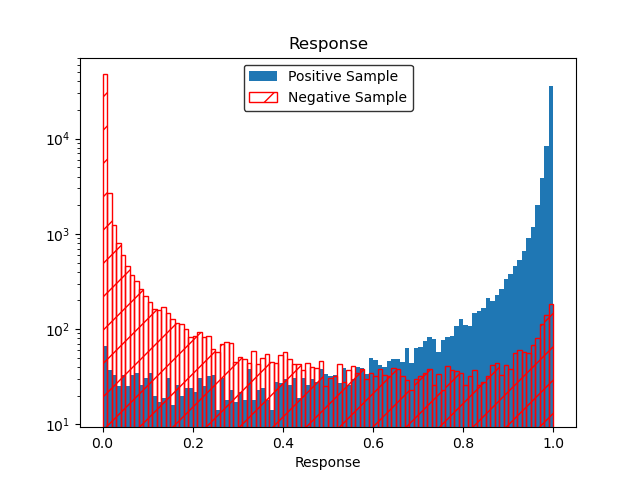}
\includegraphics[width=0.45\textwidth]{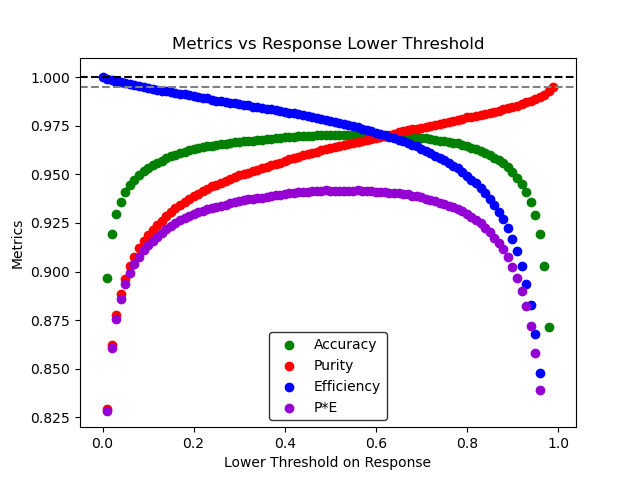}\\
\includegraphics[width=0.45\textwidth]{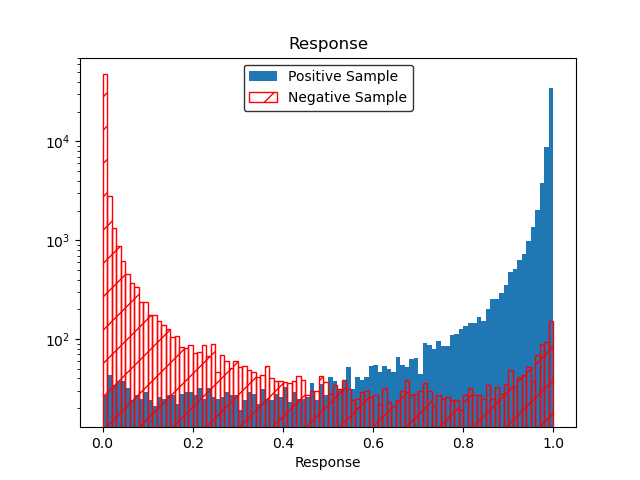}
\includegraphics[width=0.45\textwidth]{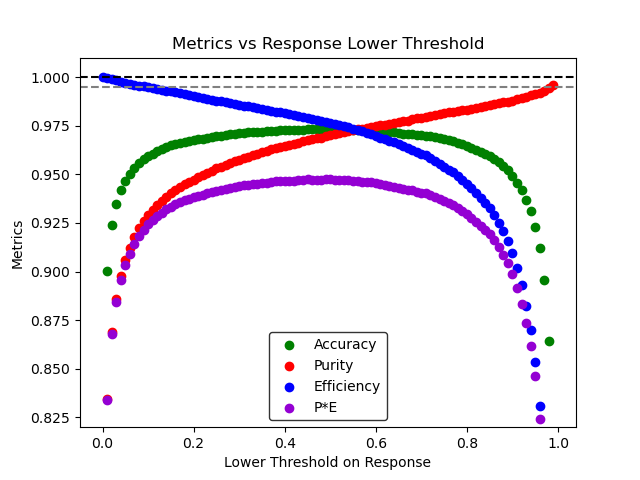}
\caption{Left: CNN Response. Right: the accuracy, efficiency and purity of the AI trigger as a function of the threshold cut on the CNN response. The black dashed line is set at 1.0, whilst the grey dashed line is set at 0.995. The top row shows the results of the trigger trained on raw hits in the DC. The bottom row shows the results of the trigger trained on de-noised tracks. For both rows, the data set used to measure the metrics has background merged so as to mimic a data set taken at 90nA.}
\label{fig:Response}
\end{figure}

\paragraph{}
The output of a CNN, and of multivariate classifiers in general, is given as the probability that an event is taken from the positive sample. This is called the response, and a perfect classifier would assign a response of 1 to all positive events and a response of 0 to all negative events. As shown in figure ~\ref{fig:Response}, for an imperfect classifier the response for both positive and negative events lies along the entire [0,1] range. The decision to select or not an event is then reduced to a lower threshold cut on the response, the value of which is chosen based on the performance of the CNN at given thresholds. For both the trigger trained on raw hits in the DC or trained on de-noised tracks, figure ~\ref{fig:Response} shows how the accuracy, efficiency and purity vary as a function of this cut on the response, with table ~\ref{table:RespMet} giving examples of the different metrics for various cuts on the response. The AI trigger should match the traditional CLAS12 electron trigger efficiency of 99.5\%, so the suggested cut on the response would here be taken at 0.08, for a purity of 91\% to 92\% depending on the training. Both the trigger trained on raw DC hits and the trigger trained on de-noised DC tracks achieved similar performance, with the trigger applied to de-noised tracks only marginally outperforming the other. For the remainder of this article it is useful to bear in mind that the aim of the AI trigger is to achieve as high a purity as possible, whilst keeping the efficiency above a certain threshold.

\begin{table}[h!]
\centering
\begin{tabular}{ |c|c|c|c||c|c|c| } 
\hline
& \multicolumn{3}{c||}{Trained on Raw DC Hits} & \multicolumn{3}{c|}{Trained on De-noised DC Tracks}\\
\multicolumn{1}{|c|}{Threshold on Response} & \multicolumn{1}{c}{Purity} & \multicolumn{1}{c}{Efficiency} & \multicolumn{1}{c||}{Accuracy} & \multicolumn{1}{c}{Purity} & \multicolumn{1}{c}{Efficiency} & \multicolumn{1}{c|}{Accuracy} \\ 
 \hline
 \hline
0.01 & 83.0 \% & 99.9 \% & 89.7 \% & 83.4 \% & 99.9 \% & 90.1 \% \\
\hline
0.04 & 88.8 \% & 99.7 \% & 93.6 \% & 89.8 \% & 99.8 \% & 94.2 \% \\
\hline
0.08 & 91.2 \% & 99.5 \% & 95.0 \% & 92.2 \% & 99.5 \% & 95.6 \%\\
\hline

\end{tabular}
\caption{The purity, efficiency and accuracy at different cuts on the response first for the trigger trained on raw hits in the DC, then for the trigger trained on denoised tracks in the DC. For both, the data set used to measure the metrics has background merged so as to mimic a data set taken at 90nA.}
\label{table:RespMet}
\end{table}

\begin{figure}[ht!]
\centering 
\includegraphics[width=0.45\textwidth]{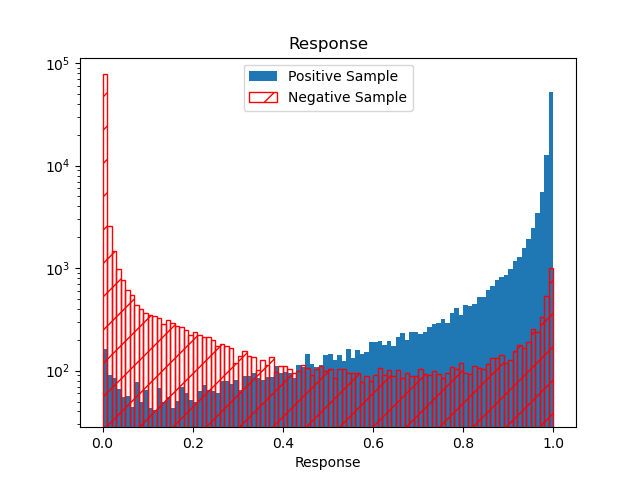}
\includegraphics[width=0.45\textwidth]{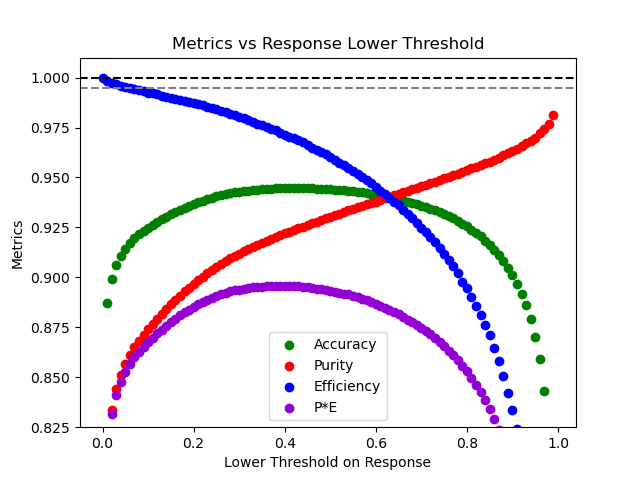}
\caption{Left: CNN Response. Right: the accuracy, efficiency and purity of the AI trigger as a function of the threshold cut on the CNN response.  The black dashed line is set at 1.0, whilst the grey dashed line is set at 0.995. Both plots demonstrate the results of the AI trigger applied to a data set taken in different experimental conditions than those on which the trigger was trained.}
\label{fig:ResponseRGM}
\end{figure}

\paragraph{}
In figure ~\ref{fig:ResponseRGM} and table ~\ref{table:RespMetRGM} we show the same metrics when applying the trigger trained on de-noised tracks in the DC to a data set taken in completely different experimental conditions, notably by decreasing the beam energy to 5.986 GeV and changing the target to a liquid deuterium target ($LD_{2}$). In this case, we apply the trigger to the clean DC tracks with no background merging as an approximation to de-noised tracks. Both figure ~\ref{fig:ResponseRGM} and table ~\ref{table:RespMetRGM} show that despite significant differences in experimental conditions the trigger maintained a reasonable performance, which is important given that at the start of a new experiment there would not be any data available to re-train the trigger. As such, this means that we can initially re-use triggers for different experiments up until a sufficient amount of data is taken to re-train the trigger.

\begin{table}[h!]
\centering
\begin{tabular}{ |c|c|c|c| } 
\hline
Threshold on Response & Purity & Efficiency & Accuracy \\ 
 \hline
 \hline
0.01 & 81.6 \% & 99.8 \% & 88.7 \% \\
\hline
0.03 & 84.4 \% & 99.7 \% & 90.6 \% \\
\hline
0.05 & 85.7 \% & 99.5 \% & 91.4 \% \\
\hline
\end{tabular}
\caption{The purity, efficiency and accuracy at different cuts on the response for the AI trigger applied to a data set taken in different experimental conditions than the trigger was trained on.}
\label{table:RespMetRGM}
\end{table}

\section{Further Testing}
\begin{figure}[ht!]
\centering  
\includegraphics[width=0.6\textwidth]{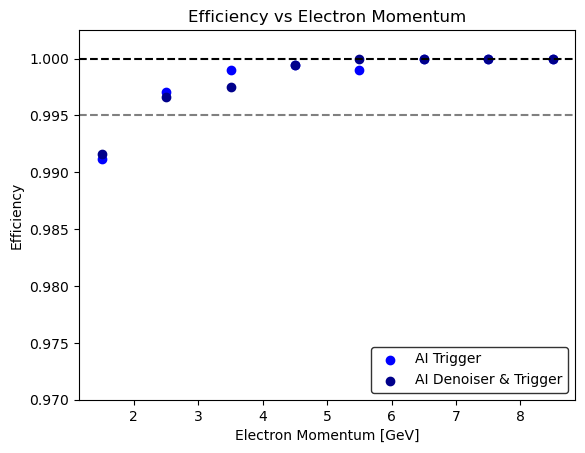}
\caption{The efficiency of the AI triggers for 1 GeV momentum bins. The data set used to measure the metrics has background merged so as to mimic a data set taken at 90nA. The threshold on the response is set at 0.08 for both triggers. The black dashed line is set a 1, and the grey dashed line is set at 0.995 as the efficiency of the current CLAS12 electron trigger is above 99.5\% in all momentum bins.}
\label{fig:MetMom}
\end{figure}

\paragraph{}
As shown in \cite{C12Trigger}, the traditional CLAS12 electron trigger is above 99.5\% over the entire momentum range. The efficiency of the AI trigger was therefore tested on data sorted into 1 GeV bins of electron momentum, as shown in figure ~\ref{fig:MetMom}, demonstrating good efficiency for all values of momentum, for both the triggers trained on noisy or de-noised DC tracks. Note that the threshold on the response could be decreased for the first momentum bin to increase the efficiency, at the cost of a decreased purity. We did not study the purity of the trigger as a function of momentum due to the fact that multiple tracks or spurious hits in a given sector makes it impossible to bin the negative sample in momentum. To calculate the efficiency we can easily bin the positive sample by making bins in electron momentum but there's an ambiguity that would confuse the evaluation of the purity as a function of momentum when deciding which track to use to bin the negative sample.

\begin{figure}[ht!]
\centering  
\includegraphics[width=0.45\textwidth]{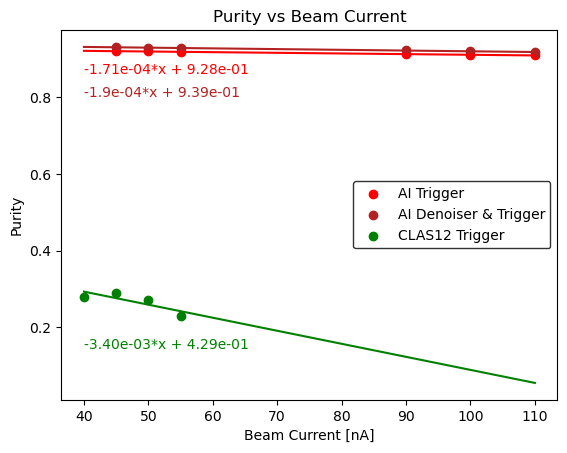}
\includegraphics[width=0.45\textwidth]{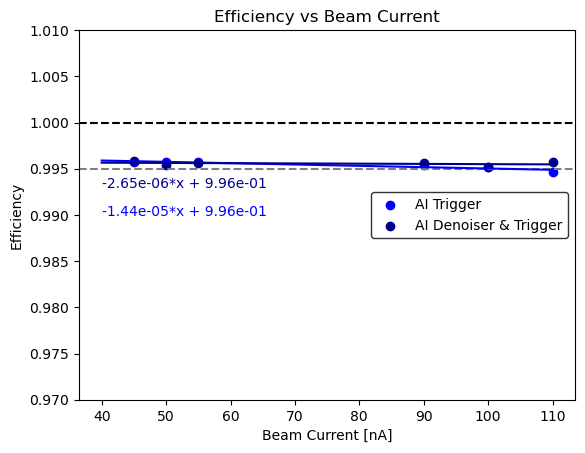}
\caption{Left: The purity of the AI (red) and traditional CLAS12 (green) triggers as a function of beam current. Right: The efficiency of the AI triggers as a function of beam currents. For both AI triggers the cut on the response was placed at 0.08 in both plots.}
\label{fig:MetBeamC}
\end{figure}

\paragraph{}
As show in table ~\ref{table:Purity} the purity of the traditional CLAS12 trigger varies with beam current, and we therefore repeated those measurements with the AI trigger trained on noisy or de-noised DC tracks. Figure ~\ref{fig:MetBeamC} shows that the AI Trigger is stable with beam current whereas the traditional CLAS12 trigger purity clearly decreases as a function of beam current. The efficiency of the AI Trigger is also stable with beam current and plotted in figure ~\ref{fig:MetBeamC}. 

\paragraph{}
We can calculate the data reduction (DR) achieved by the AI trigger relative to the traditional CLAS12 trigger as:

\begin{equation} DR=E_{AI}|(P_{AI}-P_{CLAS12})| \end{equation}

\begin{figure}[ht!]
\centering  
\includegraphics[width=0.6\textwidth]{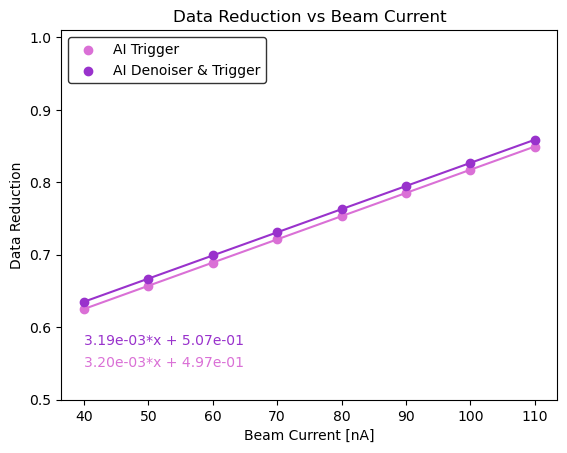}
\caption{The data reduction achieved by the AI trigger relative to the CLAS12 trigger.}
\label{fig:DataReduc}
\end{figure}

\paragraph{}
for $E$ and $P$ the efficiency and purity of the two triggers. The impact of luminosity on the trigger purity is an important consideration given future upgrades at JLab aiming to increase the luminosity. As shown in figure~\ref{fig:DataReduc} the AI trigger could considerably reduce the amount of recorded data whilst improving the purity of these data sets, which in turn will reduce costs of storage and post processing times.

\section{Software and Implementation}
\paragraph{}
As mentioned above, the AI trigger was trained and tested in python using the tensorflow library. However, to fit in with standard CLAS12 software the trigger will be deployed in Java, using the Deeplearning4j library \cite{DL4J}. The software implementation for the AI trigger has already been written and is accessible here \cite{code}. The python code used to train and test the trigger can be found in the \emph{Training} directory, with the Java code used to deploy the trigger is in the \emph{/src/main/java/org/jlab/trigger} directory of the github repository. The \emph{InputDataStream} interface is used to parse data on which to call the trigger whilst the \emph{TriggerProcessor} interface is used to apply the trigger to data parsed from the \emph{InputDataStream}. A \emph{Tester} class is provided which can operate as a standalone to read, parse and apply the trigger to a given data set before performing the tests described in this article. The general idea for deployment is that during data taking the information from different detectors will already be collated by the CLAS12 Event Builder \cite{C12Software} and passed to the Event Transfer ring, which will provide a data stream to the AI trigger. The trigger will then return the response of the CNN which can then be used to make a trigger decision. 

\begin{figure}[ht!]
\centering  
\includegraphics[width=0.6\textwidth]{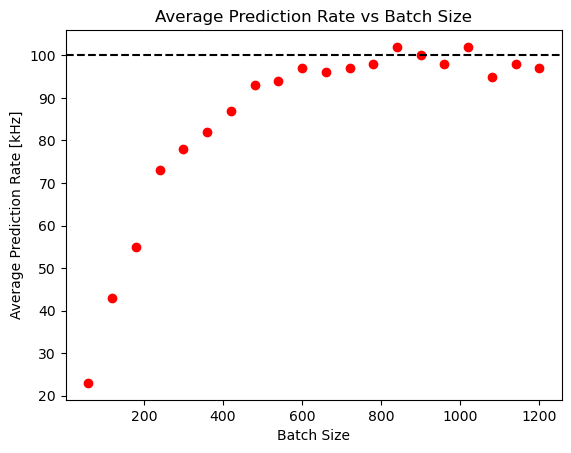}
\caption{The prediction rate of the AI trigger as a function of the batch size, for predictions grouped into batches. The prediction rate was measured using the software available at \cite{code}. Quoted here are the average rates from 100 measurements. Given that for each event the trigger must make a prediction for each of the forward detector's six sectors, the event rate for a single GPU can be roughly estimated as the prediction rate divided by six.}
\label{fig:Rates}
\end{figure}

\paragraph{}
We also tested the performance of the trigger to ensure that it can be deployed during online data taking without leading to a decrease in event rate. CLAS12 ran at a rate of 20 kHz in the past with plans to increase this in the near future. When applying the trigger to a given data set, the predictions can be split into batches with varying sizes. The prediction rate of the AI trigger trained and applied on noisy or de-noised tracks was measured for different batch sizes on a GPU with a Nvidia GeForce RTX 2080 Ti graphics card with 11 GB GDDR6 RAM and 4352 CUDA cores. Simple tests in python showed that whilst the trigger itself was able to reach a prediction rate of 100 kHz, de-noising the DC tracks requires two neural networks, which considerably slows down the prediction rate to around 40 kHz. As CLAS12 is segmented into six sectors and the trigger must make a prediction for each individual sector separately, a single GPU would only be able to reach an event rate of around 16 kHz for the AI trigger only, whilst adding the de-noiser decreases this to around 7 kHz. Given that we plan to deploy the trigger in Java, we also measured the prediction rate for the single trigger without the de-noiser using the software described above. As shown in figure ~\ref{fig:Rates} the prediction rate increases as the number of events to which the trigger is applied increases, before plateauing around 100 kHz for a batch size of 600. Furthermore, it was found that running the same process on several GPUs in parallel is slower than running several processes separately each on their own GPU.

\paragraph{}
The proposed deployment of the AI trigger would split the data stream to several GPUs working in parallel with the trigger being called on a set of events rather than individual events. With six GPUs predicting on batches of 600 events, the trigger can reach a prediction rate of 97 kHz which is more than sufficient to keep up with the data acquisition rate. A final important consideration is that, for this body of work, the trigger was tested on data that was already recorded with the CLAS electron trigger. As such, to achieve a similar performance to what was shown here, we suggest deploying the AI trigger in conjunction with the preexisting CLAS electron trigger.

\section{Discussion}
Triggers with high efficiency and purity are absolutely critical for high energy physics experiments as these tend to produce vast quantities of data, some of which may not be relevant to the given experiment. Here we have investigated the use of convolutional neural networks as an electron trigger for the CLAS12 detector. The AI trigger was trained on hits in the CLAS12 drift chambers and on the energy deposited in the strips of the CLAS12 calorimeters. Overall, the trigger was shown to improve the purity compared to the traditional CLAS12 trigger without decreasing the efficiency. This is especially true when increasing the luminosity, given that the AI trigger data reduction relative to the traditional trigger improves at a rate of 0.32\% per nA. We also compared the performance of the trigger when trained and applied on noisy or de-noised drift chamber data, resulting in the de-noised trigger achieving a marginally higher purity for a fixed efficiency but having a much decreased prediction rate due to the requirement of adding a separate neural network. The results measured for the traditional trigger and AI trigger (without de-noiser) are summarised in table ~\ref{table:Res}, with the purity for an efficiency of 99.5\% previously shown in table ~\ref{table:RespMet}, the functional form of the efficiency, purity and data reduction as a function of beam current shown in figures ~\ref{fig:MetBeamC} and ~\ref{fig:DataReduc}, and the prediction rates shown in figure ~\ref{fig:Rates}.

\begin{table}[h!]
\centering
\begin{tabular}{ |c|c|c|c|c|c| } 

 \hline \xrowht[()]{7pt}
 \cellcolor{white} Trigger & \cellcolor{white} Purity & \cellcolor{white} Efficiency (\%)  & \cellcolor{white} Purity (\%) & \cellcolor{white} Data reduction (\%) & \cellcolor{white} Prediction Rate\\
 \xrowht[()]{7pt} \cellcolor{white} & \cellcolor{white} & \cellcolor{white} vs Beam Current (nA) & \cellcolor{white} vs Beam Current (nA) & \cellcolor{white} vs Beam Current (nA) & \\
 \hline \xrowht[()]{15pt}
 \cellcolor{blue!40} AI& \cellcolor{blue!30} 91.2\% & \cellcolor{blue!40} -0.001 $x$ + 0.996 & \cellcolor{blue!30} -0.017 $x$ + 0.928 & \cellcolor{blue!40} 0.32 $x$ + 0.497 & \cellcolor{blue!30} 97 kHz  \\
 \hline \xrowht[()]{15pt}
\cellcolor{ForestGreen!40} CLAS12& \cellcolor{ForestGreen!30} 29 \% & \cellcolor{ForestGreen!40} 99.5\% & \cellcolor{ForestGreen!30} -0.34 $x$ + 0.429 & \cellcolor{ForestGreen!40} N/A & \cellcolor{ForestGreen!30} N/A \\
 \hline
\end{tabular}
\caption{For both the AI (blue) and traditional (green) triggers, the purity at an efficiency of 99.5\%, the functional form of the efficiency, purity and data reduction as a function of beam current and the prediction rate on a batch size of 600 predictions.}
\label{table:Res}
\end{table}

\paragraph{}
Another crucial requirement for a triggering system is that it must keep up with the rate of data taking. Tests made on a standard consumer GPU showed that for sufficiently large batch sizes the trigger event rate is capable of keeping up with the CLAS12 data acquisition rates, as shown in figure ~\ref{fig:Rates} and summarised in table ~\ref{table:Res} for a batch of 600 predictions. This corresponds to 100 events given that the trigger must make a prediction for each of the 6 sectors of the forward detector. Deploying the trigger on several separate GPUs would also enable faster processing and allow to keep up with high data acquisition rates. We also recommend calling the AI trigger in conjunction with the preexisting CLAS12 electron trigger to achieve a similar performance to what is shown in table ~\ref{table:Res}. The software to deploy the trigger is already available and ready for use.

\paragraph{}
Overall, the method described here promises to greatly improve on traditional-particle-identification based triggers, as it can efficiently identify the required particles whilst improving the background rejection to achieve a higher purity. This approach to triggering may prove critical given that future experiments, and possible upgrades such as higher luminosity running at JLab, aim to take increasingly larger quantities of data, at increasingly higher rates where the AI trigger outperforms the traditional trigger. Improving the triggering system will be key to reducing storing costs and post processing times for these new experiments.

\paragraph{}
As highlighted for the proposed EIC Comprehensive Chromodynamics Experiment (ECCE) detector at the up coming Electron Ion Collider (EIC), online data selection is key for monitoring purposes and near-real time analysis \cite{ECCE}. Experiments based at both JLab or the EIC will benefit from accurately identifying electrons whilst keeping up with the data taking rate. Given that most physics channels investigated at either facility will contain at least an electron, identifying electrons will allow to flag events for online monitoring or analysis, such as online calibrations and measurements of physics observables. Shorter post processing times are also key to achieving physics publication objectives. As we demonstrated here the AI trigger is adaptable to new running conditions. This flexibility means that good event selection is achievable from the start of an experiment, with shorter post processing times early on in the experiment enabling early offline monitoring, faster calibrations and eventually publication of results.

\section*{Acknowledgements}

\paragraph{}
This material is based upon work supported by the U.S. Department of Energy, Office of Science, Office of Nuclear Physics under contract DE-AC05-06OR23177 and by the U.K. Science and Technology Facilities Council under grants ST/P004458/1 and ST/V00106X/1. We would also like to thank the CLAS Collaboration for providing the data used for this body of work.

\newpage

\end{document}